\documentclass[aps,10pt,american,prb,twocolumn,nofootinbib,groupedaddress, superscriptaddress,amsmath,amssymb,longbibliography]{revtex4-1}

\usepackage{graphicx}
\usepackage{soul}
\graphicspath{{figs/}} 

\usepackage{dcolumn}
\usepackage{bm}
\usepackage{hyperref}
\usepackage{color}
\usepackage{enumitem}

\definecolor{BLACK}{gray}{0}
\definecolor{WHITE}{gray}{1}
\definecolor{RED}{rgb}{1,0,0}
\definecolor{GREEN}{rgb}{0,1,0}
\definecolor{BLUE}{rgb}{0,0,1}
\definecolor{CYAN}{cmyk}{1,0,0,0}
\definecolor{MAGENTA}{cmyk}{0,1,0,0}
\definecolor{YELLOW}{cmyk}{0,0,1,0}
\definecolor{armygreen}{rgb}{0.55, 0.73, 0.0}

\newcommand{\bs}[1]{{\color{black} #1}}

\newcommand{\ourbone}{|b_1|=45.04 \pm 0.18}

\begin{document}

\title{Numerical evidence for marginal scaling at the integer quantum Hall transition}

\author{Elizabeth J. Dresselhaus}
\affiliation{Department of Physics, University of California, Berkeley, California 94720, USA}

\author{Bj\"orn Sbierski}
\affiliation{Department of Physics, University of California, Berkeley, California 94720, USA}

\author{Ilya A. Gruzberg}
\affiliation{Ohio State University, Department of Physics, 191 West Woodruff Ave, Columbus OH, 43210, USA}

\date{\today}

\begin{abstract}
The integer quantum Hall transition (IQHT) is one of the most mysterious members of the family of Anderson transitions. Since the 1980s, the scaling behavior near the IQHT has been vigorously studied in experiments and numerical simulations. Despite all efforts, it is notoriously difficult to pin down the precise values of critical exponents, which seem to vary with model details and thus challenge the principle of universality. Recently, M. Zirnbauer\citep{Zirnbauer2019} [Nucl. Phys. B \textbf{941}, 458 (2019)] has conjectured a conformal field theory for the transition, in which linear terms in the beta-functions vanish, leading to a very slow flow in the fixed point's vicinity which we term marginal scaling. In this work, we provide numerical evidence for such a scenario by using extensive simulations of various network models of the IQHT at unprecedented length scales.
\bs{At criticality, we show that the finite-size scaling of the disorder averaged longitudinal Landauer conductance is consistent with its recently predicted fixed-point value and a third-order expansion of RG beta functions. In the future, our numerical findings can
be checked with analytical results from the conformal field theory.}
Away from criticality we describe a mechanism that could account for the emergence of an \emph{effective} critical exponents $\nu_\mathrm{eff}$, which is necessarily dependent on the parameters of the model. We further support this idea by numerical determination of $\nu_\mathrm{eff}$ in suitably chosen models.
\end{abstract}

\maketitle

\section{\label{sec:Introduction}Introduction}

A two-dimensional electron gas subject to a strong perpendicular magnetic field exhibits the integer quantum Hall effect. It is usually described in a non-interacting approximation where the number of filled Landau levels determines the (dimensionless) quantized Hall conductivity $\sigma_{xy}$. Disorder is essential as it broadens the otherwise flat Landau bands and localizes eigenstates on a scale $\xi$, so that beyond this scale the longitudinal conductivity vanishes, $\sigma_{xx} = 0$. This holds except when the energy $E$ (or field) is tuned to a critical value $E_c$ where $\xi$ diverges. The associated integer quantum Hall
transition (IQHT)\citep{Huckestein1995} belongs to the family of Anderson transitions \citep{EversMirlin:review} and is believed to be governed by a conformally-invariant fixed point in the parameter space that includes $\sigma_{xx}$ and $\sigma_{xy}$.

For a long time the commonly accepted paradigm of the IQHT fixed point \citep{Pruisken1988} was that of a conventional critical point with renormalization group (RG) beta functions whose expansions in the vicinity of the fixed point contain linear terms. In this case the RG flow equations for the deviations of the longitudinal and Hall conductivities from their fixed-point values
\begin{align}
\delta_{-}(l) &\equiv \sigma_{xx}(l) - \sigma_{xx}^{\ast},
&
\delta_{+}(l) &\equiv\sigma_{xy}(l) - \sigma_{xy}^{\ast},
\end{align}
take the form
\begin{align}
\frac{d\delta_{-}}{dl} &= y \delta_{-}+...,
&
\frac{d\delta_{+}}{dl} &= \nu^{-1}\delta_{+}+...\,\text{.}
\label{eq:beta_xy_linear}
\end{align}
Here $l$ is the logarithmic RG scale, and the fixed-point values of the conductivities, which also serve as coupling constants in the field theory, are
$\sigma_{xy}^{\ast}=1/2$ and $\sigma_{xx}^{\ast} \simeq 0.6$ where the latter is not known precisely.\citep{Schweitzer2005} Here and in the following, units of $e^2/h$ are implicit for $\sigma_{xx}$ and $\sigma_{xy}$. The ellipses denote higher order terms in $\delta_\pm$ that are usually neglected close to the fixed point $\delta_\pm=0$.

Let us note in passing that in a finite system characterized by length $L$ the RG scale $l$ cannot exceed $\ln (L/L_0)$. The initial $L_0$ (for which $l=0$) is the scale beyond which a continuum field-theory description becomes valid. In numerical simulations of discrete models one has to carefully choose $L_0$ ensuring that the system is sufficiently close to the fixed point, and the RG equations with expanded beta functions are valid. Upon choosing an appropriate $L_0$ we can extract universal data in the scaling regime between $L_0$ and $L$.

The critical exponents $\nu>0$ and $y<0$ determine the scaling behavior of observables in the vicinity of the fixed point, e.g.~the power-law divergence of the localization length $\xi(E)\sim |E-E_{c}|^{-\nu}$. In light of the notorious difficulty with analytical approaches to the IQHT, this relation is at the heart of a long history of numerical finite-size scaling studies, mostly employing the Chalker-Coddington (CC) network model.\citep{Chalker-Percolation-1988, Kramer-Random-2005, Obuse-Boundary-2008, Evers-Multifractality-2008, Slevin-Critical-2009, Obuse-Conformal-2010, Amado-Numerical-2011, Fulga2011a, Obuse-Finite-2012, Slevin-Finite-2012, Nuding-Localization-2015}
These works report $\nu = 2.56 \textendash 2.62$ but the leading irrelevant exponent $|y| \simeq 0.4$ is surprisingly small and comes with large error bars.
Consistent value of $\nu$ were also reported in a stroboscopic model of the IQHT,\citep{Dahlhaus-Quantum-2011} as well as in a recent lattice model simulations.\citep{Puschmann-Integer-2019, Puschmann-Edge-state-2021, Puschmann-Green's-2021-arXiv}

\begin{figure}[t]
\centering
\includegraphics[width=\linewidth]{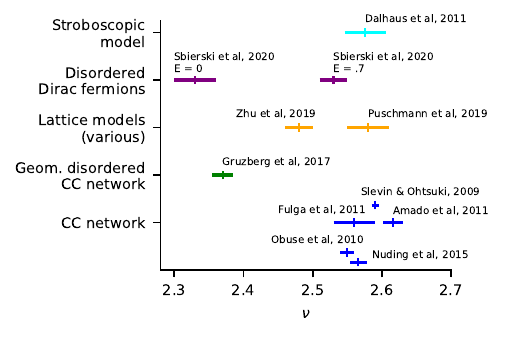}
\caption{Estimates of relevant exponent $\nu$ from various numerical studies carried between 2009 and present show statistically significant discrepancies between microscopic models in the IQHT universality class. \citep{Slevin-Critical-2009, Obuse-Conformal-2010, Amado-Numerical-2011, Fulga2011a, Dahlhaus-Quantum-2011, Obuse-Finite-2012, Slevin-Finite-2012, Nuding-Localization-2015, Gruzberg2017, Klumper2019, Puschmann-Integer-2019, Zhu-Localization-length-2019, Sbierski2020c, Puschmann-Edge-state-2021, Puschmann-Green's-2021-arXiv} Studies before 2009 did not take into account finite-size corrections to scaling, whose importance in elucidating the true critical behavior was revealed by Slevin and Ohtsuki in 2009.\citep{Slevin-Critical-2009}
\label{fig:history}}
\end{figure}

In contrast, Zhu {\it et al.}\citep{Zhu-Localization-length-2019} reported a slightly different but incompatible value $\nu=2.46 \textendash 2.50$ obtained from scaling the total number of conducting states in both lattice and continuum models projected to the lowest Landau level. Even larger deviations were reported in a structurally disordered version of the CC model\citep{Gruzberg2017, Klumper2019} where exponents as low as $\nu\approx2.37(2)$ were observed. In a recent study of models of  disordered Dirac fermions, a collaboration involving the present authors obtained $\nu \approx 2.33(3)$--$2.53(2)$ depending on the energy. \cite{Sbierski2020c} Such disordered Dirac fermions were conjectured before to be in the IQHT universality class.\citep{Ludwig1994} We summarize post-2009 results for $\nu$ in Fig. \ref{fig:history}.

To sum up, the IQHT sets itself apart from other Anderson transitions in two ways:
(i) A significant apparent variability of numerical estimates of $\nu$ across different models assumed to be in the same universality class; (ii) A very small (and possibly vanishing\citep{Amado-Numerical-2011, Nuding-Localization-2015, Puschmann-Integer-2019}) leading irrelevant exponent $|y|$.

Although (i) might be rationalized by finite size effects or the occurrence of novel universality classes, there is a more radical and intriguing alternative explanation: What if the conventional paradigm of a critical fixed point with linear beta functions does not apply to the IQHT? In Ref.~\onlinecite{Zirnbauer2019}, Zirnbauer proposed a concrete conformal field theory (CFT) for the IQHT, which, in striking contrast to all other proposals comes without relevant or irrelevant perturbations. All physically allowed perturbations turn out to be {\it marginal}, implying $\nu^{-1} = 0$ and $y = 0$. The theory moreover predicts the specific fixed-point value
\begin{align}
\sigma_{xx}^\ast = \frac{2}{\pi} \approx 0.6366
\label{sigma-fp}
\end{align}
for the longitudinal conductivity, which is related to one of the coupling constants in the theory. These results were partially based on earlier developments. \citep{Bondesan-Pure-2014, Bondesan-Gaussian-2017}

In this work, we explore the consequences of such a {\it marginal scaling} scenario on the level of the RG flow equations, and present numerical evidence for its validity. We set the stage by discussing the form of the sub-leading terms on the right-hand sides of the flow equations (\ref{eq:beta_xy_linear}) once the linear terms vanish (Sec.~\ref{sec:flow}). Along the critical line $\delta_+=0$, the equation for $\delta_-$ can be solved analytically. The result is a logarithmically slow flow of $\sigma_{xx}(L)$ towards its fixed point value $\sigma_{xx}^\ast = 2/\pi$, governed by a single universal number \bs{awaiting prediction from the CFT.\cite{Zirnbauer-Marginal-CFT-perturbations-2021}}

Using simulations of the well-established CC network model \cite{Chalker-Percolation-1988} and its much less studied two-channel generalization, \cite{Lee-Chalker-PRL1994, Lee-Chalker-PRB1994} (described in Sec.~\ref{sec:models}), we confirm the marginal scaling prediction in Sec.~\ref{sec:sigma} \bs{and give a quantiative estimate for the universal coefficient described above}. It is important that, unlike all previously studied models, the two-channel network model approaches the fixed-point conductivity from above, $\sigma_{xx}(L) > \sigma_{xx}^\ast$ as system size increases towards the thermodynamic limit.

Tuning away from criticality in Sec.~\ref{sec:mimicry}, we demonstrate how the marginal flow equations can mimic the conventional scaling with an {\it effective} exponent $\nu_\text{eff}$, offering a new perspective on the variability of numerically determined $\nu$ discussed above. If this mechanism is indeed realized at the IQHT, why is the so-far-observed variation of $\nu_\text{eff}$ only in the few percent range? Do models with a drastically different value of $\nu_\text{eff}$ exist? To answer these questions, we first show that $\nu_\text{eff}$ is controlled by the longitudinal conductivity $\sigma_{xx}$ in the fixed point's vicinity, which is numerically close in all standard models for which high-accuracy estimates of $\nu$ have been obtained. Crucially, as stated above, the two channel network model is an exception and indeed realizes $\nu_\text{eff} \simeq 3 \textendash 4$, consistent with the above mechanism (Sec.~\ref{sec:nu_eff}). We present conclusions and directions for future work in Sec.~\ref{sec:conclusion}.

\begin{figure}[t]
\centering
\includegraphics{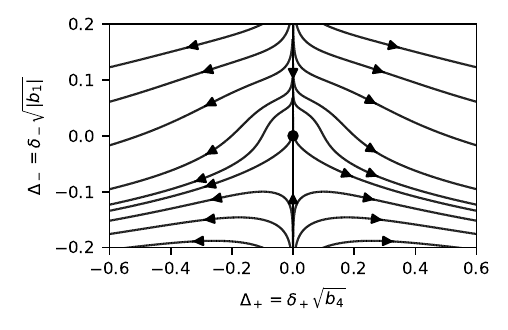}
\caption{Flow diagram based on Eqs. (\ref{eq:beta_xx}) and (\ref{eq:beta_xy}). The flow is depicted for the rescaled RG variables $\Delta_\pm$ (see axis labels) and some choice of the parameters $b_i$. The fixed point at $(0,0)$ is denoted by a dot.
\label{fig:flow}}
\end{figure}

\section{\label{sec:flow}Marginal flow equations}

We now explore the consequences of Zirnbauer's proposal for the flow equations (\ref{eq:beta_xy_linear}) and assume $\sigma_{xx}^\ast = 2/\pi \approx 0.6366$ from now on. If the linear terms on the right-hand side vanish, higher order contributions in $\delta_\pm$ have to be taken into account. Based on symmetries of the Pruisken field theory\citep{Pruisken1984, Levine} (periodicity in $\sigma_{xy}$ and behavior under reversal of the magnetic field), Khmelnitskii argued that $\frac{d\delta_{-}}{dl}$ must be even in $\delta_{+}$ and $\frac{d\delta_{+}}{dl}$ must be odd, and proposed a global flow diagram. \cite{Khmelnitskii-Quantization-1983} To respect the topology of the flow diagram we require that the flow is always away from the axis $\delta_{+}=0$ for both signs of $\delta_-$ (no term $\delta_+\delta_-$ in $\frac{d\delta_{+}}{dl}$). Likewise, to get a fixed point which is stable along the $\delta_+=0$ axis, no term $\delta_-^2$ can appear in $\frac{d\delta_{-}}{dl}$. Thus, we arrive at the RG equations
\begin{align}
\frac{d\delta_{-}}{dl} &= b_{1} \delta_{-}^{3} + b_{2} \delta_{+}^{2}+...,
\label{eq:beta_xx}
\\
\frac{d\delta_{+}}{dl} &= b_{3} \delta_{-}^{2} \delta_{+} + b_{4} \delta_{+}^{3} + ...\,\text{.}
\label{eq:beta_xy}
\end{align}
The expected phenomenology of the IQHT requires $b_{1,2}<0$ and $b_{3,4}>0$. \bs{Despite recent analytical advances with the candidate CFT,\cite{Zirnbauer-Marginal-CFT-perturbations-2021} the parameters $b_{1,2,3,4,...}$ remain unknown.} For a certain ad-hoc choice of these parameters the flow is depicted in Fig.~\ref{fig:flow}. We note that similar flow equations (with a different $b_2$-term of the form $b_2 \delta_- \delta_+^2$) have been suggested.\citep{Zirnbauer-localisation2020-talk}

We emphasize that in the marginal scaling scenario, the values of  $\sigma_{xx}^\ast$ and the parameters $b_{1,2,3,4}$ in Eqs.~\eqref{eq:beta_xx} and \eqref{eq:beta_xy} are the universal data that take over the role of $\nu$ and $y$ in defining the IQHT universality class, both conceptually and from an applied point of view.


\section{\label{sec:models}Network models}

\begin{figure}
\centering
\includegraphics{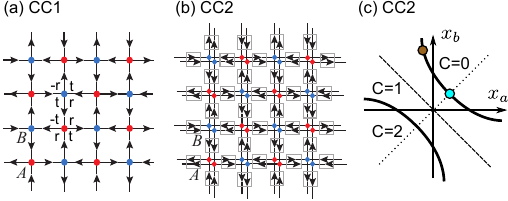}
\caption{Network models employed in this work. Panel (a) shows the standard CC model (CC1). Panel (b) shows a two-layer generalization of the CC1 termed CC2, which features layer-preserving node scattering parametrized by the tuple $(x_a,x_b)$. Panel (c) shows the schematic phase diagram of the CC2 in the $x_a$-$x_b$ plane following Ref.~[\onlinecite{Lee-Chalker-PRB1994}]. The dotted line indicates the (ensemble) symmetry under layer exchange $x_a \leftrightarrow x_b$, the dashed line represents the symmetry of the bulk phase diagram under $(x_a,x_b) \rightarrow (-x_a,-x_b)$ and $C$ denotes the number of edge states for a finite system with a specific choice of boundary termination. The colored dots correspond to the critical parameter values $(x_a,x_b) = (-0.483,3)$ (brown) and $(x_a,x_b) = (0.227,0.227)$ (cyan) identified in Sec.~\ref{sec:nu_eff}.
\label{fig:CCmodels}}
\end{figure}

For numerical simulations of the IQHT in subsequent sections we rely on network models originally introduced based on semiclassical arguments \cite{Chalker-Percolation-1988} and widely applied due to their numerical efficiency. \cite{Kramer-Random-2005} We work with the unitary symmetry class variants of these networks models, whose critical behavior is generally accepted to belong to the IQHT universality class. The standard CC model [abbreviated CC1, see Fig.~\ref{fig:CCmodels}(a)] is defined on a checkerboard lattice with inequivalent sites A,B (dots) at which the incoming chiral states on the links (arrows) are scattered quantum mechanically into two possible outgoing states with scattering amplitudes
\begin{align}
r &= \frac{1}{\sqrt{1 + e^{-2x}}},
&
t &= \frac{1}{\sqrt{1 + e^{2x}}},
\label{eq:rt}
\end{align}
controlled by the model's single parameter $x$, which encodes the probabilities for right and left turns. The disorder is realized by U$(1)$-phases $e^{i\phi_j}$ with $\phi_j\in[0,2\pi)$ associated randomly to each link $j$.

The two-layer (or two-channel) generalization of the CC1, termed CC2, features two parallel chiral channels per link, see Fig.~\ref{fig:CCmodels}(b). Without loss of generality, the scattering at the node is layer-preserving and parametrized by the tuple $(x_a,x_b)$ as above. The disorder, which causes both inter- and intra-layer scattering, is modeled by Haar-random U$(2)$ matrices acting on co-moving states on the links (boxes). This model has been introduced in Refs.~[\onlinecite{Lee-Chalker-PRL1994}, \onlinecite{Lee-Chalker-PRB1994}], where the qualitative structure of the phase diagram, reproduced schematically in Fig.~\ref{fig:CCmodels}(c), was revealed.

The bulk phase diagram of the CC$N$, $N=1,2$ can be understood by the mirror symmetry across a straight line of links, mapping a disorder realization with parameter $x$ to one in the $-x$ ensemble. Together with the fact that the CC$N$ generically has $N+1$ topologically distinct phases with $C=0,1,..,N$ edge states,\cite{Khmelnitskii-Quantization-1983} this fixes the critical point of the CC1 to $x=0$. The above mirror symmetry also gives rise to the dashed symmetry line in the phase diagram for the CC2.  The dotted symmetry line for the CC2 phase diagram in Fig.~\ref{fig:CCmodels}(c) arises from a statistical layer-exchange symmetry $x_a \leftrightarrow x_b$.

An important practical complication for the CC2 and any even $N$ is that the positions of critical lines are not fixed by any symmetry argument but have to be found numerically. We defer the description of our numerical approach to this task to Sec.~\ref{sec:nu_eff}. On the other hand, due to the two-dimensional parameter space, the CC2 offers the possibility to tune along the critical line, a feature of paramount importance to our study that is absent in the CC1.

The critical properties of the CC2, so far assumed to be in the IQHT universality class at all points along the critical line, are not known with great accuracy due to large localization lengths and the aforementioned uncertainty about the critical $(x_a,x_b)$. In  Ref.~[\onlinecite{Lee-Chalker-PRB1994}], the authors settled for a modified model with an ad-hoc weakening of inter-layer scattering and reported $\nu=2.45$ from a study of quasi-1d Lyapunov exponents at a certain point on the critical line. No error bars were given. Our results for the CC2 with full interlayer scattering presented in Sec.~\ref{sec:nu_eff} are significantly different.

All observables defined and computed in subsequent sections are based on the steady-state, four-terminal scattering matrix $S$ of large network models. Each terminal refers to the union of incoming and outgoing links along one of the four sides of a rectangular-shaped network. To numerically obtain $S$ for systems of linear size up to order $10^4$ efficiently, we use an iterative patching approach, concatenating the scattering matrices of four rectangular subsystems with size $L_x \times L_y$ into the scattering matrix of a single system of size $2L_x \times 2L_y$. We define
\begin{align}
a = L_y/L_x
\end{align}
as the aspect ratio. Note that unlike transfer matrix multiplication, the scattering matrix concatenation does not require further numerical stabilization. The main computational bottleneck limiting system sizes is the large memory required to store the iteratively obtained matrices $S$. A useful feature of this iterative approach is that systems of exponentially different sizes are generated in a single run.


\section{\label{sec:sigma}Marginal scaling at criticality}

In this section, we focus on the critical line $\delta_{+}=0$.
In this case, the right-hand side of Eq.~(\ref{eq:beta_xy}) vanishes
while Eq.~(\ref{eq:beta_xx}) for $\delta_-(l)=\sigma_{xx}(l)-\sigma_{xx}^\ast$  becomes $\frac{d\delta_{-}}{d l} = b_{1}\delta_{-}^{3}$, and we neglect higher order terms. This can be integrated up to $l = \ln (L/L_0)$:
\begin{equation}
\delta_{-}(L) = \frac{\delta_{-}(L_{0})}{\sqrt{1 + 2 |b_{1}| \delta_{-}^{2}(L_{0}) \ln\left(L/L_{0}\right)}}.
\label{eq:dm_crit}
\end{equation}
As we have already mentioned, the initial scale $L_0$ has to be chosen large enough for the expansion in Eq.~\eqref{eq:beta_xx} to apply but is arbitrary beyond that requirement. In the numerical tests of Eq.~\eqref{eq:dm_crit} that we perform in the following we identify $L$ with the minimum of the two lengths $L_x$ and $L_y$. That is,
\begin{align}
L &= \begin{cases}
       L_x, & \mbox{for } a > 1 \\ L_y, & \mbox{for } a < 1.
     \end{cases}
\end{align}
We then fix the value of $\delta_{-}(L_{0})$ from numerical data and check  if the form of the numerically obtained $\delta_{-}(L)$ for $L>L_0$ follows Eq.~\eqref{eq:dm_crit}.

An important remark is in order here. \bs{The exact CFT prediction $\sigma_{xx}^\ast = 2/\pi$ and the yet unknown value of $b_1$ that should be found from a bulk CFT describes the flow of the coupling constant of the field theory, and, a priori, is} not directly related to transport properties of a finite-size system with non-trivial geometry and specific boundary conditions at attached leads. Such transport properties need to be independently computed from Kubo formulas as certain correlators in the field theory. At present, the only available result of this type is the average (dimensionless) conductance $g_{xx}^\ast$ of a cylinder with two absorbing leads and arbitrary aspect ratio $a$, but only at the fixed point,\cite{Zirnbauer-Marginal-CFT-perturbations-2021} given by
\begin{equation}
g_{xx}^{\ast}\left(a\right) = \sigma_{xx}^\ast a \sum_{m\in\mathbb{Z}} (-1)^{m}
e^{-\pi^{2} m^{2}\sigma_{xx}^\ast a}.
\label{eq:g*}
\end{equation}
Note that for large aspect ratios $a\gg1$, the fixed-point bulk coupling constant and the conductance are simply related by $\sigma_{xx}^\ast = g_{xx}^\ast(a) /a$.

The field theory result, Eq. \eqref{eq:g*}, applies in the scaling limit $L_x, L_y \to \infty$. In contrast, our numerical simulations are restricted to finite systems. In the following, we use rectangular systems of finite size $L_x\times L_y$ where $L_y=a\cdot L_x$. We attach absorbing leads attached in the $x$ direction and periodic boundary conditions in the $y$ direction, realizing a cylinder geometry. The two-terminal scattering matrix defined in terms of the modes in the left and right leads can be found from the four-terminal scattering matrix of a $L_x\times L_y$ system by short-circuiting the transverse leads. Then the two-terminal conductance can be computed from the Landauer formula. \cite{NazarovBlanterBook} It depends on the specific disorder realization in the system and we report the associated histograms in Appendix \ref{app:histo}. In the following, we denote its disorder average by $g_{xx}(a,L_x)$.

To date, there is no field theory result for $g_{xx}$ away from the fixed point.
Therefore, to make progress for numerically accessible finite system sizes, we resort to the ad-hoc assumption that Eq.~\eqref{eq:g*} remains valid in the vicinity of the fixed point for $1\ll L<\infty$. We thus assume
\begin{equation}
g_{xx}(a,L) = \sigma_{xx}(L) a \sum_{m\in\mathbb{Z}} (-1)^{m} e^{-\pi^{2}m^{2}\sigma_{xx}(L)a},
\label{eq:gL}
\end{equation}
with $\sigma_{xx}(L) = \sigma_{xx}^\ast + \delta_-(L)$, where $\delta_-(L)$ is given by Eq.~\eqref{eq:dm_crit}. Note that Eq. \eqref{eq:gL} is at best an approximation to the correct result, since in the field theory of Ref. [\onlinecite{Zirnbauer-Marginal-CFT-perturbations-2021}] not only the coupling constants but also the current operators that enter Kubo formulas get modified away from the fixed point.

Using the network models introduced in the previous section, we numerically
assess the validity of Eq.~(\ref{eq:gL}) and determine the value of $b_{1}$. While for CC1 the critical point occurs at $x=0$, for the CC2 we tune to $(x_a,x_b)=(-0.483,3)$ [brown dot in Fig.~\ref{fig:CCmodels}(c)] which will be shown to be on the critical line below in Sec.~\ref{sec:nu_eff}. We treat the cases $a<1$ and $a>1$ separately.

\begin{figure}
\centering
\includegraphics{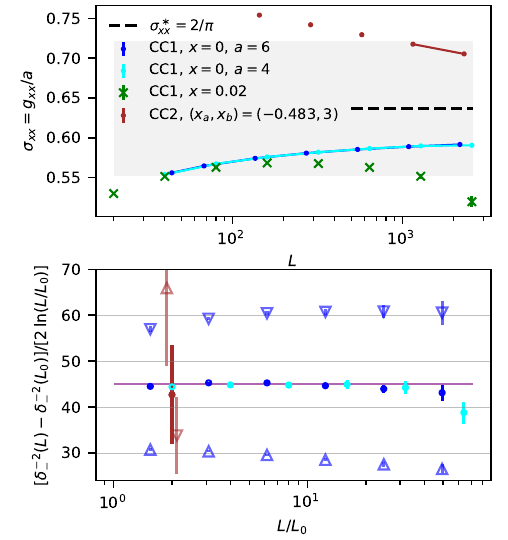}
\caption{Numerical assessment of marginal scaling of $\delta_- = \sigma_{xx} - \sigma_{xx}^{\ast}$ at criticality $\delta_+ = 0$ for large aspect ratios $a=4,6$. The top panel shows the disorder averaged conductivity $\sigma_{xx}$ (dots) obtained from the Landauer conductance of wide slabs with aspect ratio $a\gtrsim 4$ for the CC1 and CC2 network model. 
The CC1 data in the vicinity of $\sigma_{xx}^{\ast}$ (gray box) connected by solid lines follow the scaling prediction in Eq.~(\ref{eq:dm_crit_check}) with $\ourbone$ (purple line, fitted for $a=6$), see bottom panel. We use $L_0=40$ and $44$ for $a=4$ and $6$, respectively. The upward (downward) triangles denote the CC1 and CC2 data for ad-hoc variations of $\sigma_{xx}^\ast$ by $+0.01$ ($-0.006$). CC1 data (blue triangles) show a clear $L$-dependence, while the CC2 data move away from the purple line in the opposite way compared with CC1. Data for the CC1 at small finite $x=0.02$ is also included in the top panel to show that the constant $b_2$ in Eq.~(\ref{eq:beta_xx}) is negative.}
\label{fig:sigma}
\end{figure}

We start with large aspect ratios $a=4,\,6$ where to very high precision the exponential corrections in Eq.~\eqref{eq:gL} are negligible and we obtain $g_{xx}/a=\sigma_{xx}$. We find approximately Gaussian histograms for the conductances, see App.~\ref{app:histo}. This can be rationalized by the picture of a large number $\sim a$ of effectively decoupled parallel conductors with random conductances. The data in the top panel in Fig.~\ref{fig:sigma} shows a mo\-no\-to\-nously increasing (decreasing) flow of $g_{xx}/a$ for CC1 (CC2) with the conjectured critical conductivity $\sigma_{xx}^{\ast}=2/\pi$ consistently placed between the two data sets. We remark that to the best of our knowledge, this is the first numerical observation of a decreasing $\sigma_{xx}(L)$ at a IQHT, hypothesized long ago in the proposed flow diagram of Khmelnitskii.\cite{Khmelnitskii-Quantization-1983}

Due to the simple form of the conjectured Eq.~\eqref{eq:gL} for large aspect ratios, $g_{xx}/a=\sigma_{xx}$, we can attempt a direct numerical determination of the parameter $b_1$. In Fig.~\ref{fig:sigma} (bottom)
we plot the numerically obtained quantity
\begin{align}
|b_1(L)| &\equiv \frac{\delta_{-}^{-2}(L) - \delta_{-}^{-2}(L_{0})}{2\ln\left(L/L_{0}\right)},
\label{eq:dm_crit_check}
\end{align}
which, according to Eq.~(\ref{eq:dm_crit}) should be length-independent and identified with the universal number $|b_{1}|$. The only freedom is the choice of the initial system length $L_{0}$ which must be large enough such that $|\delta_{-}(L_{0})|$ is sufficiently small to apply the expansion of the flow equations in Sec.~\ref{sec:flow}. Indeed, if we consider the CC1 data points for which $|\delta_{-}(L_{0})| \leq 0.085$ (grey region, data points connected by solid lines) we confirm that $|b_1(L)|$ in Eq.~(\ref{eq:dm_crit_check}) is practically a constant over almost two decades in $L/L_0$. A least-squares fit yields (purple line)
\begin{equation}
\ourbone.
\label{eq:b1}
\end{equation}
\bs{We now investigate the stability of the so found parameter when (i) the CC2 model is considered and (ii), the value of $\sigma_{xx}^*$ is varied.}

(i) For the CC2 model at $(x_a,x_b)=(-0.483,3)$ (brown), there are only two data points in the grey region, which are nevertheless consistent with Eq.~\eqref{eq:b1}. Sliding along the CC2 critical line to the point $(x_a,x_b)=(0.227,0.227)$, we observe $\sigma_{xx}(L=1920)\simeq0.78$ way outside the scaling region (data not shown). This is consistent with the observation of very large localization length in the CC2 model made in Refs.~[\onlinecite{Lee-Chalker-PRL1994}, \onlinecite{Lee-Chalker-PRB1994}].
Sliding along the critical line in the other direction (towards larger $x_b$) did not lower $\sigma_{xx}$ appreciably.

(ii) To further test the prediction $\sigma_{xx}^* =2/\pi$, we repeated the above analysis with ad-hoc variations of $\sigma_{xx}^\ast$ by $+0.01$ ($-0.006$). This led to drastically different values of $|b_1(L)|$, see upward (downward) triangles. The resulting values for CC1 (blue triangles) exhibit a considerable dependence on $L$. Significantly, for CC2 (brown triangles) the resulting values shift in the opposite way compared with CC1. Thus, the expected universality of $b_1$ in the marginal flow scenario holds only for the precise value $\sigma_{xx}^\ast=2/\pi$, lending additional support for the prediction of Ref.~\onlinecite{Zirnbauer2019}.

Finally, in Fig.~\ref{fig:sigma} (top) we also include data for the CC1 at small, finite $x=0.02$ (green crosses) which show an initial increase toward $\sigma_{xx}^*$ for small $L$, but then curve away from the fixed-point conductivity at large $L$. This confirms the negative sign of the constant $b_2$ in Eq.~(\ref{eq:beta_xx}).

\begin{figure}
\centering
\includegraphics{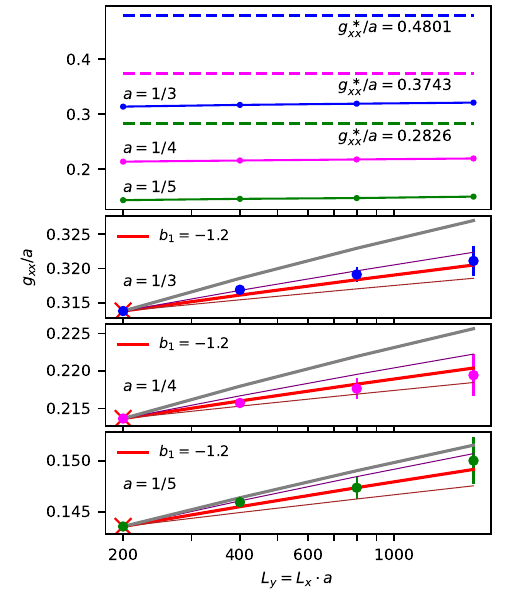}
\caption{
Numerical assessment of marginal scaling for the critical disorder averaged Landauer conductance $g_{xx}$ (dots) for small aspect ratios $a=1/3,\;1/4,\;1/5$. The data is obtained from the CC1 network model with $x=0$ and the solid lines are guides to the eye. The top panel compares the numerical finite-size results with the CFT prediction for the fixed point conductance, see Eq.~\eqref{eq:g*} (dashed lines). The three lower panels are close-ups on the numerical data (dots) comparing the slow increase of conductance from the reference scale $L_y=200$ to the postulated analytical form in Eq.~\eqref{eq:gL} \bs{using $b_1=-1.2$ (red line)}. The thin brown and purple lines denote the analytical form with a $30\%$ decreased or increased value of $b_1$, respectivley. The gray line is obtained from Eq.~\eqref{eq:gL} by neglecting the flow of $\sigma_{xx}$ in the exponent.
}
\label{fig:sigma_smallAspect}
\end{figure}

We next turn to small aspect ratios $a<1$. Fig.~\ref{fig:sigma_smallAspect} shows $g_{xx}/a$ (dots) at fixed small aspect ratios $a=1/3,\;1/4,\;1/5$ as a function of system size $L_y$. Here we limit ourselves to the CC1. The associated histograms in Appendix~\ref{app:histo} are approximately log-normal with a hard cutoff at unit conductance. This peculiar behavior is reminiscent of transport in quasi one-dimensional systems\cite{Beenakker-RMT-RMP-1997} and the associated paucity of conductances of order unity requires at least $10^4$ disorder realizations to obtain acceptable error bars on the average conductances reported in Fig.~\ref{fig:sigma_smallAspect}. The increase of $g_{xx}(a,L_x)/a$ (dots) towards the CFT fixed point values in Eq.~\eqref{eq:g*} (dashed lines) is very slow, see the bottom panels of Fig.~\ref{fig:sigma_smallAspect} for a close-up. This slow change is potentially in agreement with the marginal flow in Eq.~\eqref{eq:dm_crit}. Attempts to extract $|b_1|$ using Eq. \eqref{eq:dm_crit_check} to fit the small-$a$ data (not shown) result in values of $|b_1| \sim O(1)$, much smaller than for the large-$a$ data.

However, for $a < 1$ the exponential corrections in Eq.~\eqref{eq:gL} are significant and require a different \bs{approach to assess the validity of the marginal scaling in Eq.~\eqref{eq:dm_crit}.}
We use the value of $g_{xx}/a$ at the smallest system size $L_y=200$ (red cross) on the left hand side of Eq.~\eqref{eq:gL} to fix $\delta_-(L_{y,0}=200)$. \bs{Making the ad-hoc choice $b_1=-1.2$, we compute the expected variation of $g_{xx}/a$ for $L_y>L_{y,0}$, see the solid red line in the bottom panels in Fig.~\ref{fig:sigma_smallAspect}. We obtain excellent agreement with the numerical data within error bars for all aspect ratios studied. The gray lines are the result of a similar procedure where in Eq.~\eqref{eq:gL} only the prefactor $\sigma_{xx}(L_x)$ is flowing according to Eq.~\eqref{eq:dm_crit} (again with $b_1=-1.2$). The difference in slope emphasizes the importance of a flowing $\sigma_{xx}$ in the exponential.}

In summary, if our ad-hoc assumption \eqref{eq:gL} for the relation between the finite-size Landauer conductance $g_{xx}(a,L_x)$ and the bulk flow equation for $\delta_-(L)$ in Eq.~\eqref{eq:dm_crit} is reasonable, our numerical results for \bs{small aspect ratios $a<1$ predict $b_1=-1.2$.} \bs{A possible reason why two different $b_1$ are obtained for large or small aspect ratios $a>1$ and $a<1$, respectively, is the dominant role of absorbing boundary conditions in the first case. Absorbing boundaries strongly affect the interference of wave packets injected by the leads.}

\section{\label{sec:mimicry}Mimicry of relevant scaling from marginal RG flow}
\begin{figure}
\centering
\includegraphics{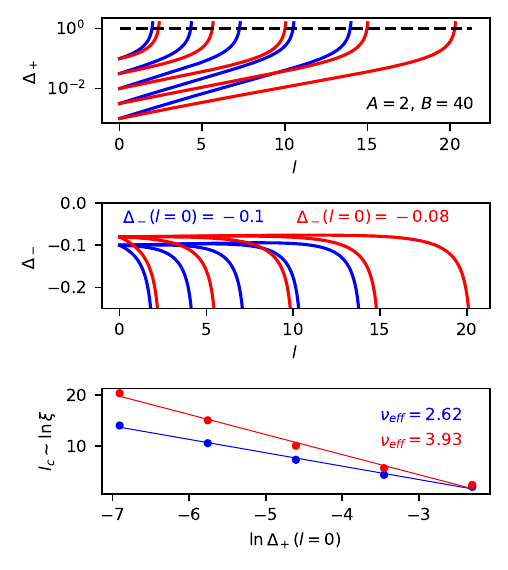}
\caption{Numerical solution of the rescaled flow equations (\ref{eq:dDelta-}) and (\ref{eq:dDelta+}) with the parameter choice $A=2$, $B=40$, $\Delta_-(0) = -0.1$ (blue) and $-0.08$ (red), respectively. Various $\Delta_+(0) = 10^{-3}...10^{-1}$ are chosen as detailed in the bottom panel. Each flow is stopped at $l=l_c$ defined as $\Delta_{+}(l_{c}) = 1$ (dashed line in top panel). The relation between the initial $\Delta_+(0)$ and $l_c$ shown in the bottom panel approximately follows $l_c \sim - \nu_\text{eff} \, \ln\Delta_{+}(0) + \text{const}$ (solid lines) with $\nu_\text{eff}$ dependent on $\Delta_{-}(0)$.}
\label{fig:mimicry}
\end{figure}

In this section, we demonstrate how the marginal flow equations in Sec.~\ref{sec:flow} can give rise to an apparent conventional scaling of the localization length, $\xi\sim|\delta_{+}(L_0)|^{-\nu_\text{eff}}$. To reduce the number of constants in the flow equations of Sec.~\ref{sec:flow}, we define rescaled variables $\Delta_\pm(l)$ and (unknown) universal numbers $A$, $B$:
\begin{align}
\Delta_{-}(l) &\equiv \sqrt{|b_1|} \, \delta_-(l),
&
\Delta_{+}(l) &\equiv \sqrt{b_4} \, \delta_+(l),
\nonumber \\
A &=\sqrt{|b_1|}\frac{|b_2|}{b_4},
&
B &= \frac{b_3}{|b_1|}.
\end{align}
In terms of these, and neglecting higher-order terms, the RG Eqs. \eqref{eq:beta_xx} and \eqref{eq:beta_xy} become
\begin{align}
\frac{d\Delta_{-}}{dl} &=
-\Delta_{-}^{3} - A\Delta_{+}^{2},\label{eq:dDelta-}\\
\frac{d\Delta_{+}}{dl} & = B\Delta_{-}^{2} \Delta_{+} + \Delta_{+}^{3}\label{eq:dDelta+}.
\end{align}
As usual, the localization length $\xi$ is defined via $\ln(\xi/L_0) \sim l_{c}$ with $l_{c}$ the RG cutoff time given by $\Delta_{+}(l_{c})=1$.

The qualitative behavior of the solutions of the flow equations \eqref{eq:dDelta-} and \eqref{eq:dDelta+} very close the fixed point can be obtained by neglecting the second terms on the right-hand side. Then $\Delta_-(l)$ flows logarithmically slow toward zero (see Eq. \eqref{eq:dm_crit} above), and the factor $B \Delta_-^2(l)$ can be approximately treated as a constant in front of $\Delta_+(l)$, as in Eq. \eqref{eq:beta_xy_linear}, resulting in
\begin{align}
\nu_\text{eff} \sim [B\Delta_{-}^{2}(l=0)]^{-1}.
\label{eq:nueff}
\end{align}
This is the mimicry of the conventional critical scaling as defined above.

Eventually, RG trajectories leave the vicinity of the fixed point and  $\Delta_+(l)$ grows large enough so that the second terms on the right-hand side in \eqref{eq:dDelta-} and \eqref{eq:dDelta+} significantly alter the global flow. The condition for the mimicry of the conventional scaling is that the RG trajectory spends a long time sufficiently close to the fixed point where the powers of $\Delta_+(l)$ higher than the linear can be neglected, and $\Delta_{-}^{2}(l)$ does not vary appreciably across this range of $l$. This is easily achieved for $A \ll 1$ and $B \gg 1$ by starting the flow with $|\Delta_{-}(0)| \ll 1$. In this case we still expect that $\nu_\text{eff} \sim [B\Delta_{-}^{2}(0)]^{-1}$. This expectation is confirmed by the exact solution of the system \eqref{eq:dDelta-} and \eqref{eq:dDelta+} in the case $A=0$, presented in Appendix \ref{app:exact}, where Eq.~\eqref{eq:nueff} is shown to hold for $B \gg 1$.

The mimicry mechanism turns out quite robust and can be confirmed by the full numerical solution of Eqs.~(\ref{eq:dDelta-}) and (\ref{eq:dDelta+}). The results of the previous section suggest that typical values of $\Delta_{-}(0)$ \bs{might be} of order $-0.1$. Employing $\Delta_{-}(0) = -0.1$ along with the ad-hoc choice $A=2$ and $B=40$ for the universal parameters, Fig.~\ref{fig:mimicry} shows a numerical solution, see blue lines. A linear approximation to the data in the bottom panel using $\ln\xi \sim -\nu_\text{eff} \ln\Delta_{+}(0) + \text{const}$ yields $\nu_\text{eff} \simeq 2.62$. The red lines denote results when the initial value $\Delta_{-}(0)$ is slightly changed to $-0.08$, in this case $\nu_\text{eff} \simeq 3.93$ emerges. Both values are slightly beyond the accuracy of the estimate \eqref{eq:nueff} ($2.44$ and $3.81$, respectively), since $B = 40$ is not quite large enough.

Larger values of $B$ would reduce the curvature in Fig.~\ref{fig:mimicry} (bottom) and thus better approximate conventional scaling, but also push towards smaller values of $\nu_\text{eff}$. The numerical value of $A$ is of minor importance for $\nu_\text{eff}$, as is demonstrated by the exact solution for $A=0$. The same is true for the precise value of the constant used in the definition of $l_c$ above, as long as it is of the order of unity. This is clear from the steep slope (and the eventual divergence) of $\Delta_+(l)$ close to $l_c$, see Fig.~\ref{fig:mimicry} (top). In addition, we have confirmed (results not shown) that the mimicry mechanism is qualitatively unchanged for the alternative flow equations proposed in Ref.~\onlinecite{Zirnbauer-localisation2020-talk}, indicating that the precise nature of the higher-order terms does not play a significant role.
We stress, however, that for all known numerical models of the IQHT, the quantitative validity of the above truncation of the flow equations leading to \eqref{eq:dDelta-} and \eqref{eq:dDelta+}, is questionable. The reason is that in a specific model it is impossible to tune $|\Delta_{-}| = \sqrt{b_1} \, |\sigma_{xx} - 2/\pi|$ to arbitrarily small values. Thus, it may be necessary to include higher order terms in Eqs.~\eqref{eq:beta_xx} and \eqref{eq:beta_xy} to capture the flow of $\Delta_\pm$ from their starting values. These effects and their quantitative influence on the mimicry of conventional scaling are left for future work.

In summary, we have shown how the conjectured marginal flow equations \eqref{eq:beta_xx} and \eqref{eq:beta_xy} can approximately mimic conventional scaling with an effective critical exponent $\nu_\text{eff}$. There are two qualitative conclusions that could serve as hallmark signatures of the marginal scaling scenario:
\begin{enumerate}[label=(\roman*)]
  \item The dependence of the effective critical exponent $\nu_\text{eff}$ on $\delta_{-}(L_0)$ and through it on the chosen model and its parameter values, c.f.~Fig.~\ref{fig:mimicry}. Although the relation is likely more complicated than the simple estimate $\nu_\text{eff} \sim \delta_{-}^{-2}(0)$, we should expect sizeable variations of $\nu_\text{eff}$ between models for the IQHT if they realize different $\delta_{-}(0)$.
  \item The relation $\xi \sim |\delta_{+}(0)|^{-\nu_\text{eff}}$ is only approximately fulfilled. Indeed, Fig.~\ref{fig:mimicry} (bottom) reveals a small residual curvature, that can be captured by the introduction of a $\delta_{+}(0)$-dependent critical exponent. Anticipating the relation $\delta_{+}(0) \sim \delta_x \equiv x-x_c$ for the network models, we thus use the ansatz
\begin{equation}
\xi = \lambda |x-x_c|^{-\nu_\text{eff}(\delta_x)}.
\label{eq:nueff(x)}
\end{equation}
Here $\lambda$ is a non-universal parameter with the dimension of length.
\end{enumerate}

In the next section, we investigate both signatures (i) and (ii) with exact numerical simulations of the network models. While we confirm a model and parameter dependent critical exponent $\nu_\text{eff}$, we can only put an upper bound on a putative $\delta_x$ dependence of $\nu_\text{eff}(\delta_x)$ in the CC1.

\section{\label{sec:nu_eff}Numerical demonstration of variable $\nu_\text{eff}$ in models of the IQHT}
\begin{figure*}
\centering
\includegraphics[width = \textwidth]{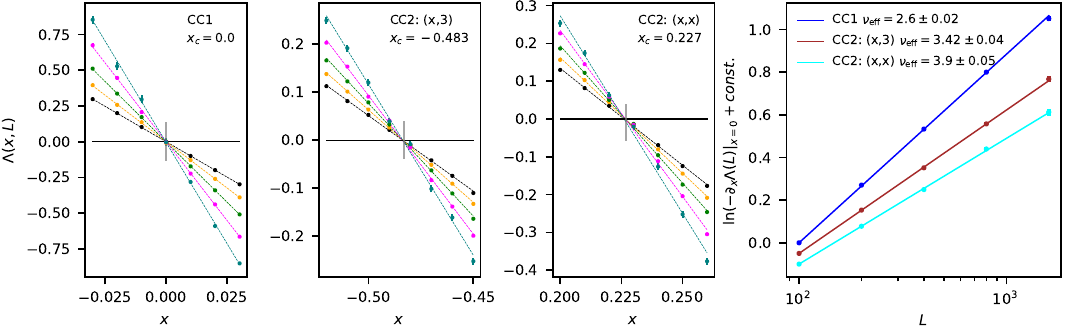}
\caption{The left three panels show the scaling variable $\Lambda(x,L)$ for $x$ crossing the critical point $x_c$ and square systems of size $L = 100, 200, 400, 800, 1600$. The right panel shows the slopes $-\partial_x \Lambda(x,L)$ obtained from linear fits to each individual system size $L$. They follow power-laws $\sim L^{1/\nu_\text{eff}}$ with various $\nu_\text{eff}$ indicated in the legend. The data points are offset in the vertical direction for clarity.
\label{fig:Lambda}
}
\end{figure*}

We now tune our numerical models away from their critical points $x=x_c$ to study the divergence of the localization length close to criticality which we assume to be described by Eq.~\eqref{eq:nueff(x)}. We adopt the scattering-matrix based observable $\Lambda$ initially proposed by Fulga \emph{et al.}, Ref.~[\onlinecite{Fulga2011a}], and recently employed to study scaling for the 2d Dirac model.\cite{Sbierski2020c} This observable is qualitatively similar to the scaling variable $\delta_+ = \sigma_{xy} - 1/2$ in that they both change sign at the critical point. This property provides  a simple and precise method to determine $x_c$ for models like the CC2 where the critical point is not fixed by symmetry.

In this section, we fix the aspect ratio to $a=1$ and wrap the $L\times L$ system in one direction, called the transverse direction, to form a cylinder. To define $\Lambda(x,L)$ for the cylinder, we attach a lead extended over the full width of the system to one of the open ends of the cylinder. Then we consider the reflection matrix $r(\phi)$ of the lead as a function of the phase $\phi$ of twisted boundary conditions in the transverse direction, or equivalently the value of an Ahronov-Bohm flux piercing the cylinder. For a given disorder realization, as we change $x$, the critical point $x=x_c$ occurs when there exists a $\phi$ such that $r(\phi)$ has a zero eigenvalue and thus $\det r(\phi)=0$. This follows directly from the definition of a reflection-matrix based topological invariant in the unitary symmetry class.\cite{Fulga2011a}

To obtain $\Lambda$, we generalize the transverse boundary condition from a phase factor to an arbitrary complex number $z \in \mathbb{C}$. In terms of scattering states $\psi_{i,j}$ defined on all links of the network indexed by pairs of integers $(i,j)$, the generalized boundary condition reads: $\psi_{i,j=L-1} = z\, \psi_{i,j=0}$ for all $i = 0, 1, ..., L-1$. This additional freedom allows for solutions $z_0$ of $\det r(z_0) = 0$ to exist even away from criticality $x \neq x_c$. However, generalized zeros of this kind occur away from the unit circle, $|z_{0}|\neq1$. To measure the distance to criticality, consider $\mathrm{log}|z_{0}|$ for the $z_0$ closest to the unit circle. This quantity indeed changes sign at $x=x_c$ and features a Gaussian histogram in the ensemble of disorder realizations.\cite{Sbierski2020c} Finally, the scaling observable is defined as $\Lambda=\overline{\ln|z_{0}|}$ where the overline denotes disorder average.

We proceed with the conventional single-parameter scaling hypothesis \cite{Cardy-Scaling-1996} stating that a dimensionless observable like $\Lambda$ depends not on the system size $L$ and the dimensionless parameter $x$ separately, but as $\Lambda(x,L) = \Lambda(L/\xi(x))$, with the localization length $\xi$ from Eq.~\eqref{eq:nueff(x)}. Requiring that $\Lambda(x,L)$ is analytic in $\delta_x\equiv x-x_c$ and using the property $\Lambda(x=x_c,L)=0$, we expand
\begin{equation}
\Lambda(x,L) = \alpha_{1} \delta_x \Big(\frac{L}{\lambda} \Big)^{\nu_\text{eff}^{-1}(\delta_x)}
+ \alpha_{2} \delta_x^2 \Big(\frac{L}{\lambda} \Big)^{2 \nu_\text{eff}^{-1}(\delta_x)} + ...
\label{eq:expansion}
\end{equation}

We now turn to the demonstration of the hallmark signature (i) from the previous section: a sizeable model- and parameter dependence of $\nu_\text{eff}$. We anticipate the $\delta_x$-dependence of $\nu_\text{eff}$ [point (ii)] to be a comparatively small effect which we neglect for now and revisit at the end of this section. In Fig.~\ref{fig:Lambda}, we report $\Lambda$ for CC1 and CC2 network models of size $L=100,200,400,800,1600$, aspect ratio $a=1$ and several $x$ around $x_c$. While $x_c=0$ for the CC1 by symmetry, for the CC2 we focus on two line-cuts in the phase diagram of Fig.~\ref{fig:CCmodels}(c), $(x_a,x_b)=(x,3)$ and $(x,x)$, respectively. We take the range of $x$-values small enough so that the higher order terms in Eq.~\eqref{eq:expansion} do not contribute, see Fig.~\ref{fig:Lambda}. For each $L$, we perform a fit of $\Lambda(x,L)$ linear in $x$ (dotted line) and extract its zero-crossing. Remarkably, these crossing points agree for all $L$ within an accuracy better than $\Delta x_c=0.001$, indicating the practical absence of  corrections to scaling. The average value $x_c$ is reported in Fig.~\ref{fig:Lambda} and was used for the study of the critical longitudinal conductivity in Sec.~\ref{sec:sigma}.

The CC2 is known to have localization lengths large compared to CC1,\cite{Lee-Chalker-PRB1994} reflected by a larger $\lambda$. In our study, this leads to drastically smaller slopes for the CC2 when compared to the CC1 at the same $L$, see Fig.~\ref{fig:Lambda}. In practice, this requires hundreds of thousands of disorder realizations to achieve an acceptable ratio between data point separation and error bars.

To extract $\nu_\text{eff}$, we employ the scaling prediction $\Lambda(x,L)/\delta_x \sim L^{1/\nu_\text{eff}}$, valid for small enough $\delta_x$ and $L$, c.f.~Eq.~\eqref{eq:expansion}. We approximate the left hand side by the slopes of the linear fits mentioned above, and plot the slopes in the right panel of Fig.~\ref{fig:Lambda} where we extract $\nu_\text{eff}$. For the CC1, our result $\nu_\text{eff} = 2.60(2)$ is compatible with the generally accepted value, and in particular with the value $2.56(3)$ obtained by Fulga \emph{et al.}~\cite{Fulga2011a} using the same method and model but for a very different aspect ratio $a=5$. Our main result, however, is the demonstration of very different exponents for the CC2: $\nu_\text{eff} = 3.42(4)$ and $\nu_\text{eff} = 3.90(5)$ at the critical points $(-0.483,3)$ and $(0.227,0.227)$, respectively.

Finally, we turn to the second hallmark signature (ii), according to which the exponent $\nu_\text{eff}$ in Eq.~\eqref{eq:nueff(x)} should depend on $\delta_x$. We stress that a check of this prediction requires an analysis of numerical data at several \emph{fixed} values of $x$, which is usually not attempted in the existing literature. Such an analysis crucially relies on
our ability to represent $\Lambda(x,L)$ as the expansion \eqref{eq:expansion} with a non-trivial exponent $\nu_\text{eff}(\delta_x)$.

\begin{figure}
\centering
\includegraphics{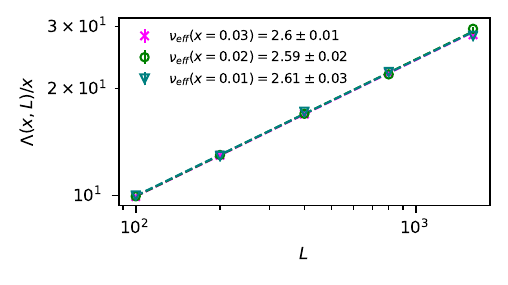}
\caption{Scaling plot for $\Lambda(x,L)/x$ at fixed $x$ computed from the CC1 at aspect ratio $a = 1$ for $L = 100, 200, 400, 800, 1600$. Data for $x = 0.01, 0.02, 0.03$ agree within error bars and the power law fit [Eq.~\eqref{eq:nueff(x)check}] (dashed lines) yields compatible values for $\nu_\text{eff}(x)$ for all three $x$.
\label{fig:LamOverX}
}
\end{figure}

We focus on the CC1 for its numerical convenience, exactly known $x_c=0$ (implying $\delta_x = x$), and odd-in-$\delta_x$ expansion [implying $\alpha_2 = 0$ in Eq.~\eqref{eq:expansion}],
and select the data for $x=0.01,0.02,0.03$ from Fig.~\ref{fig:Lambda}, left panel. In Fig.~\ref{fig:LamOverX} we show that the anticipated scaling relation
\begin{equation}
\ln \frac{\Lambda(L,x)}{x} = \ln \alpha_1 + \frac{1}{\nu_\text{eff}(x)}
\ln \frac{L}{\lambda}
\label{eq:nueff(x)check}
\end{equation}
holds with $\nu_\text{eff}(x)$ values that agree within error bars for all chosen $x$ and are, moreover, consistent with the value of $\nu_\text{eff}$ obtained above using the slope of $\Lambda$ extracted from multiple values of $x$. Decreasing $x$ further was found not to be suitable due to enhanced statistical error bars, while larger $x$ would require including higher-order corrections in Eq. \eqref{eq:expansion} and the corresponding modifications to
Eq.~\eqref{eq:nueff(x)check}.

Let us note here that in principle we can always trade the dependence of the effective exponent $\nu_\text{eff}$ on $\delta_x$ for the dependence of $\nu_\text{eff}$ on the system size $L$. Indeed, any functional relation $\xi(x)$ can be inverted to produce a function $x(\xi)$, see Eq. \eqref{eq:x-of-xi} or \eqref{eq:x-L} in Appendix \ref{app:exact} as an example. Then a scaling function $f[L/\xi(x)]$ can be traded for another scaling function $F[x/x(L)]$. Once this replacement is done, the function $x(L)$ can always be written as $x_0 L^{-1/\nu_\text{eff}(L)}$, giving the scaling function of the form $F[x L^{1/\nu_\text{eff}(L)}]$ with an $L$-dependent effective exponent $\nu_\text{eff}(L)$. For the conventional power-law scaling, when $\xi \propto |x|^{-\nu}$, all these transformations are rather trivial and well known: they correspond to the replacement of $f(L|x|^\nu)$ by $F(x L^{1/\nu})$.

In our case the empirical observation that $\nu_\text{eff}$ does not depend on $x$  is completely consistent with the visible absence of any curvature in the right panel in Fig. \ref{fig:Lambda}, as well as in Fig. \ref{fig:LamOverX}.

In summary, using a variation of $x$ by a factor of three, we were not able to positively identify the proposed $\delta_x$ dependence of $\nu_\text{eff}$ anticipated in the marginal scaling scenario. However, it is not obvious to us that this should be taken as a serious argument against the validity of the latter. Indeed, the curvature of the data in the bottom panel of Fig.~\ref{fig:mimicry} sensitively depends on the unknown universal parameters in the flow equations. Detecting a small curvature corresponding to a weak $x$-dependence of $\nu_\text{eff}$ might very well require varying $x$ by one or two orders of magnitude. Likewise, detecting an $L$-dependent $\nu_\text{eff}$ might require much larger system sizes $L$. Both options are currently beyond the capability of our numerical approach.

\section{\label{sec:conclusion}Conclusion and outlook}
In this work, we scrutinized the consequences of Zirnbauer's marginal scaling scenario\cite{Zirnbauer2019} recently conjectured for the IQHT. We proposed and analyzed a beyond-linear-order expansion of the RG beta functions for the longitudinal and Hall conductivities taking into account the known topology of the flow diagram. At criticality, our numerical simulations of the one- and two-channel network models confirm that the resulting RG equation indeed describes the flow of the critical longitudinal conductance with the system size $L$. The RG flow depends on a universal number $|b_1|$ which we determined numerically to depend only on the \bs{aspect ratio $a$, $|b_1|\simeq1.2$ for small aspect ratio $a\ll 1$ and $\ourbone$ for large aspect ratio $a\gg 1$. We attribute this surprising geometry dependence to the subtle role of the absorbing boundary conditions imposed by the ideal leads.
As the bulk quantity $|b_1|$ is a unique and universal number for the marginal scaling scenario of the IQHT, further analytical work is clearly needed to shed light on this issue.}

Moving away from the critical point, we showed by a proof-of-principle numerical solution of the conjectured flow equations how an \emph{effective} localization length exponent $\nu_\text{eff}$ can arise from marginal scaling where formally $\nu^{-1}=0$. The effective exponent depends on the critical conductivity at short distances, albeit the detailed functional relation probably requires an account of higher-order terms in the flow equations, which is left for future work. Interestingly, it appears that the selection of models for the IQHT studied in the literature so far conspired to have very close critical conductivities and, accordingly, only moderate (less than 10\%) variations of $\nu_\text{eff}$ from the value $\nu_\text{CC1}=2.6$ accepted for the CC1. These variations were mostly blamed to uncertainties in the fitting procedure or finite-size effects.

To reach larger variation of $\nu_\text{eff}$, models with drastically increased tuning capabilities are required. One example is the 2d Dirac model, \cite{Ludwig1994} where energy and disorder strength can be varied independently. As shown recently,\cite{Sbierski2020c} $\nu_\text{eff} \simeq 2.33$ was found at half-filling, $E=0$, and a certain disorder strength, the largest deviation from $\nu_\text{CC1}=2.6$ reported to date. As shown in the present work, the two-channel network model CC2 whose critical conductivity decays with increasing system sizes produces very different results for $\nu_\text{CC2}$ between $3$ and $4$ and thus provides strong evidence for the validity of the marginal scaling picture.

We emphasize that according to the conjectured flow equations, a larger absolute value of $\delta_-(L_0)=\sigma_{xx}(L_0)-2/\pi$, i.e.~a larger distance of short-length longitudinal conductivity to the fixed point value $2/\pi$ should lead to a smaller $\nu_\text{eff}$. Naively, comparing our results for the CC1 and CC2 shows the opposite trend. This can be rationalized by the intrinsic difficulty in defining the length scale $L_0$ beyond which a beta-function description of $\sigma(L)$ becomes possible at all. And even for $L \gtrsim L_0$, an expansion of the beta functions beyond the order considered above might be required.

Very recently, the IQHT was also studied in the framework of dissipation-induced topological states\cite{Beck2020} and the result $\nu \simeq 3$ was interpreted as a signature of a novel non-equilibrium universality class. Further work will be necessary to see if this interpretation is correct or if dissipative systems might also fit in the marginal scaling framework advocated here.

\begin{acknowledgments}
We acknowledge useful discussions with John Chalker, Ferdinand Evers, Matthew Foster, Igor Gornyi and Martin Zirnbauer. Computations were performed at the Ohio Supercomputer Center and the Lawrencium cluster at Lawrence Berkeley National Lab.
E.J.D was supported by the NSF Graduate  Research Fellowship  Program,  NSF DGE 1752814. B.S. acknowledges financial support by the German National Academy of Sciences Leopoldina through grant LPDS 2018-12.

\end{acknowledgments}

\vspace{3mm}

\appendix

\section{Histograms of two-terminal Landauer conductance}
\label{app:histo}

\begin{figure}
\centering
\includegraphics{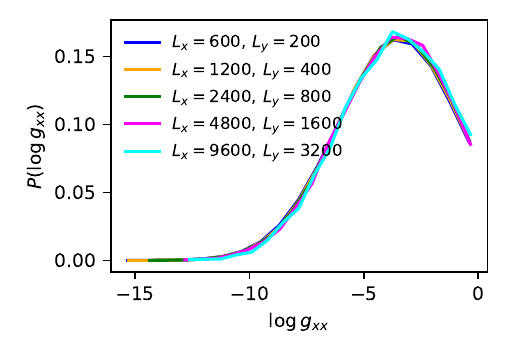}
\includegraphics{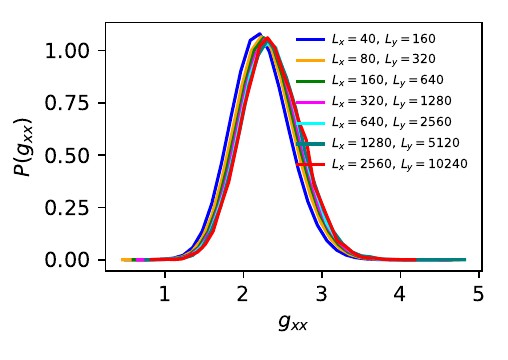}
\caption{
Histograms for Landauer conductance $g_{xx}(a)$ of a critical CC1 model ($x=0$) at aspect ratio $a=1/3$ (top) and $a=4$ (bottom).
}
\label{fig:histo}
\end{figure}

In Fig.~\ref{fig:histo}, we show the histograms for Landauer conductance $g_{xx}(a)$ of a critical CC1 model ($x=0$) at aspect ratio $a=1/3$ (top) and $a=4$ (bottom). In the small aspect ratio case, we observe $g_{xx}\leq 1$ with an approximately log-normal distribution. The large aspect ratio shows no such cutoff and follows an approximately gaussian distribution of $g_{xx}$ with a mean proportional to $a$. The well known histogram for a square sample\citep{Kramer-Random-2005} $(a=1)$ with a kink of the distribution function at $g_{xx}=1$ and a small tail of $g_{xx}>1$ can be thought of as a interpolation between the cases for large and small $a$ reported here.

\section{Exact solution of the flow equations with $A=0$}
\label{app:exact}

\begin{figure}
\centering
\includegraphics{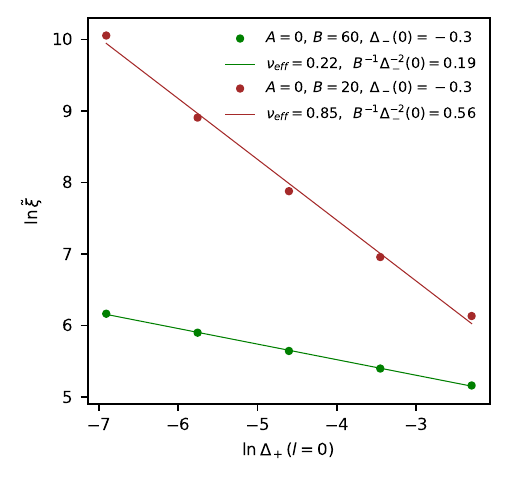}
\caption{Analytical solution for $\ln \tilde{\xi}$ given in Eq. \eqref{eq:tilde-xi}. This figure is similar to the bottom panel in Fig.~\ref{fig:mimicry}. As everywhere in this appendix $A=0$, while we chose $B=20$ (green) and $B=60$ (brown) for the two data sets. The dots represent of $\ln \tilde{\xi}$ computed from Eq.~\eqref{eq:tilde-xi} for the same values of $\Delta_+(0)$ as in the bottom panel in Fig.~\ref{fig:mimicry}, and $\Delta_-(0) = -0.3$.}
\label{fig:mimicry-analytic}
\end{figure}

When we neglect the second term in the right-hand side of Eq. \eqref{eq:dDelta-}, the flow equations become
\begin{align}
\frac{d\Delta_{-}}{dl} & = -\Delta_{-}^{3},
&
\frac{d\Delta_{+}}{dl} &= B\Delta_{-}^{2} \Delta_{+} +\Delta_{+}^{3}
\label{eq:dDelta+A0}.
\end{align}
This system is exactly solvable. The first equation \eqref{eq:dDelta+A0} is solved in the same way as in Section \ref{sec:sigma}:
\begin{align}
\Delta_-(l) = \Delta_-(0) (1 + 2 \Delta_-^2(0) l)^{-1/2}.
\end{align}
Then the second equation \eqref{eq:dDelta+A0} becomes a linear equation in terms of the variable $X(l) \equiv \Delta_+^{-2}(l)$:
\begin{align}
\frac{d X}{dl} &= - 2B \Delta_-^2(l) X - 2.
\end{align}
This equation has the solution
\begin{align}
X(l) &= e^{-F(l)}\bigg(X_0 - 2 \int_{0}^{l} e^{F(l')} dl' \bigg),
\nonumber \\
X_0 &\equiv \Delta_+^{-2}(0),
\qquad
F(l) = 2B \int_{0}^{l} \Delta_-^2(l') dl'.
\end{align}
Performing the integrals, we obtain
\begin{widetext}
\begin{align}
\frac{1}{\Delta_+^2(l)}
&= \frac{[1 + 2 \Delta_-^2(0) l]^{-B}}{\Delta_+^2(0)}
- \frac{[1 + 2 \Delta_-^2(0) l] - [1 + 2 \Delta_-^2(0) l]^{-B}}{(B+1) \Delta_-^2(0)}.
\label{eq:Delta+A0-solution}
\end{align}
Rewritten in terms of $\Delta_-(l)$ this equation describes the integral curves of the system \eqref{eq:dDelta+A0} in the $\Delta_+ \textendash \Delta_-$ plane:
\begin{align}
\frac{\Delta_+^2(0)}{\Delta_+^2(l)}
&= \Big(\frac{\Delta_-^2(0)}{\Delta_-^2(l)}\Big)^{-B}
- \frac{\Delta_+^2(0)}{(B+1) \Delta_-^2(0)}
\bigg[\frac{\Delta_-^2(0)}{\Delta_-^2(l)} - \Big(\frac{\Delta_-^2(0)}{\Delta_-^2(l)}\Big)^{-B} \bigg].
\end{align}
\end{widetext}
Notice that the resulting flow is even in $\Delta_-(l)$, which is a consequence of setting $A=0$. This property is lost for the full system \eqref{eq:dDelta-}, \eqref{eq:dDelta+}.

The RG flow described by Eq.~\eqref{eq:Delta+A0-solution} begins at $l=0$ with a very large left-hand side $\Delta_+^{-2}(0) \gg 1$. As $l$ grows, the first term on the right-hand side decreases, while the second (negative) term increases in magnitude. When we reach the localization length, $l_c = \ln (\xi/L_0)$, the left-hand side becomes one, and the resulting equation can be numerically solved for $\ln (\xi/L_0)$. However, for our current analysis it is better to adopt a slightly different definition of $\ln (\xi/L_0)$ as the RG time for which $\Delta_+(\tilde{l}_c) = \infty$. With this modification the terms on the right-hand side cancel each other when $l = \tilde{l}_c$. This gives a relation between $\Delta_+(0)$ and $\tilde{\xi}$:
\begin{align}
\Delta_+^2(0) &= \frac{(B+1)\Delta_-^2(0)}{[1 + 2 \Delta_-^2(0) \ln (\tilde{\xi}/L_0)]^{B+1} - 1},
\label{eq:x-of-xi}
\end{align}
which can be inverted to give
\begin{align}
\ln \frac{\tilde{\xi}}{L_0} = \frac{1}{2 \Delta_-^2(0)} \bigg( \Big[ (B+1) \frac{\Delta_-^2(0)}{\Delta_+^2(0)} + 1 \Big]^{\frac{1}{B+1}} - 1 \bigg).
\label{eq:tilde-xi}
\end{align}
This quantity is shown in Fig.~\ref{fig:mimicry-analytic} for $B=20$ and $B=60$, together with linear fits resulting in effective exponents $\nu_\text{eff}$ whose values are given in the legend.

Let us estimate $\nu_\text{eff}$ analytically. If we start the flow sufficiently close to the critical line, $\Delta_+(0) \ll 1$, the first term in the square brackets in Eq. \eqref{eq:tilde-xi} dominates, and we get approximately
\begin{align}
\ln \frac{\tilde{\xi}}{L_0} \approx \frac{[(B+1) \Delta_-^2(0)]^{\frac{1}{B+1}}}{2 \Delta_-^2(0)} [\Delta_+(0)]^{-\frac{2}{B+1}} - \frac{1}{2 \Delta_-^2(0)}.
\end{align}
Now comes the mimicry. If $B \gg 1$, the factor with $\Delta_+(0)$ has a very small exponent, and can be approximated by
\begin{align}
[\Delta_+(0)]^{-\frac{2}{B+1}} & = \exp \Big(-\frac{2}{B+1} \ln\Delta_+(0)\Big)
\nonumber \\&
\approx 1 -\frac{2}{B+1} \ln\Delta_+(0).
\end{align}
Notice that this approximation only works for $|\ln\Delta_+(0)| \ll B$, that is, sufficiently {\it far} from criticality. When legitimate, this approximation leads to a linear relation between $\ln (\tilde{\xi}/L_0)$ and $\ln\Delta_+(0)$ with the coefficient
\begin{align}
\nu_\text{eff} = [(B+1) \Delta_-^2(0)]^{-\frac{B}{B+1}}
\approx [B \Delta_-^2(0)]^{-1}.
\label{eq:nu-eff-crude}
\end{align}
This analytical estimate is given in the legend of Fig. \ref{fig:mimicry-analytic} for the same values of $B$ and $\Delta_-(0)$ as the actual values of $\ln \tilde{\xi}$. We see that the agreement between the analytical estimates and the results of linear fits becomes reasonable only for unrealistically large values of $B$ ($B \gtrsim 50$ for our choice of $\Delta_-(0) = -0.3$ and the range of $\Delta_+(0)$).

The effective exponent in Eq. \eqref{eq:nu-eff-crude} can also be obtained directly from the $L$ scaling. Indeed, following the discussion at the end of Section \ref{sec:nu_eff}, we replace $\tilde{\xi}$ in Eq. \eqref{eq:x-of-xi} by $L$ and write
\begin{align}
x^2(L) = \frac{(B+1)\Delta_-^2(0)}{[1 + 2 \Delta_-^2(0) \ln (L/L_0)]^{B+1} - 1}.
\label{eq:x-L}
\end{align}
Sufficiently close to criticality and for not too large $L/L_0$ we can replace
\begin{align}
\Big(1 + 2 \Delta_-^2(0) \ln \frac{L}{L_0}\Big)^{B+1} &\approx \exp \Big( 2(B+1)\Delta_-^2(0) \ln \frac{L}{L_0} \Big)
\nonumber \\
&= \Big(\frac{L}{L_0} \Big)^{2(B+1)\Delta_-^2(0)}.
\end{align}
Then it immediately follows that for large $B$ we obtain $x(L) \propto L^{-1/\nu_\text{eff}}$ with $\nu_\text{eff} = [(B + 1)\Delta_-^2(0)]^{-1}$.

\bibliography{library}

\end{document}